\documentclass[letterpaper]{article} 
\usepackage{aaai24}  
\usepackage{times}  
\usepackage{helvet}  
\usepackage{courier}  
\usepackage[hyphens]{url}  
\usepackage{graphicx} 
\usepackage{natbib}  
\usepackage{caption} 
\usepackage{algorithm}
\usepackage{algorithmic}
\usepackage{amsmath,amsfonts}
\usepackage{booktabs}  
\usepackage{multirow}
\newtheorem{proposition}{Proposition}

\usepackage{newfloat}
\usepackage{listings}
\DeclareCaptionStyle{ruled}{labelfont=normalfont,labelsep=colon,strut=off} 
\floatstyle{ruled}
\newfloat{listing}{tb}{lst}{}
\floatname{listing}{Listing}
\title{Robust Communicative Multi-Agent Reinforcement Learning with Active Defense}
\author{
    Lebin Yu, Yunbo Qiu, Quanming Yao, Yuan Shen, Xudong Zhang and Jian Wang \thanks{Corresponding Author}
}
\affiliations{
    Department of Electronic Engineering, BNRist, Tsinghua University, Beijing 100084, China \\
    \{yulb19, qyb18\}@mails.tsinghua.edu.cn, quanmingyao@gmail.com, \{shenyuan\_ee, zhangxd, jianwang\}@tsinghua.edu.cn
}

\usepackage{bibentry}

\begin{document}

\maketitle

\begin{abstract}
Communication in multi-agent reinforcement learning (MARL) has been proven to effectively promote cooperation among agents recently. Since communication in real-world scenarios is vulnerable to noises and adversarial attacks, it is crucial to develop robust communicative MARL technique. However, existing research in this domain has predominantly focused on passive defense strategies, where agents receive all messages equally, making it hard to balance performance and robustness. We propose an active defense strategy, where agents automatically reduce the impact of potentially harmful messages on the final decision. There are two challenges to implement this strategy, that are defining unreliable messages and adjusting the unreliable messages' impact on the final decision properly. To address them, we design an Active Defense Multi-Agent Communication framework (ADMAC), which estimates the reliability of received messages and adjusts their impact on the final decision accordingly with the help of a decomposable decision structure. The superiority of ADMAC over existing methods is validated by experiments in three communication-critical tasks under four types of attacks.
\end{abstract}

\section{Introduction}
In recent years, multi-agent reinforcement learning (MARL) has made remarkable strides in enhancing cooperative robot tasks and distributed control domains, exemplified by traffic lights control \cite{chu2019multi} and robots navigation \cite{han2020cooperative}. Given the inherent partial observability in multi-agent tasks, several researchers have explored the integration of communication mechanisms to facilitate information exchange among agents \cite{ma2021learning}. 


Nevertheless, the utilization of multi-agent communication in real-world applications introduces certain challenges. In such scenarios, agents rely on wireless communication channels to exchange messages, which are susceptible to various sources of noises and interference. These perturbations on messages, especially malicious ones, can severely reduce the performance of multi-agent systems, even though agents get accurate observations of the environment \cite{sun2023certifiably}. Hence, it becomes imperative to address these issues and devise robust communicative MARL mechanisms.

\begin{figure*}[ht]
  \centering
  \includegraphics[width=0.95\textwidth]{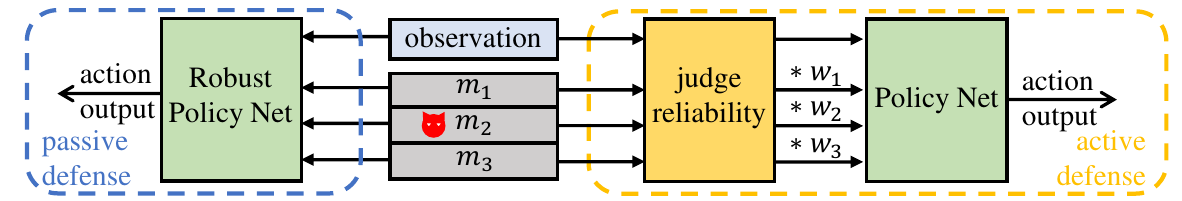}
  \caption{The figure shows the difference between active defense and passive defense. Passive defense takes in harmful messages along with useful ones equally, and try to make robust decisions. This indiscriminate reception makes it difficult to optimize robustness and performance simultaneously. In comparison, active defense assesses the reliability of incoming messages with local information first and reduces the weight of potentially malicious messages. For example, $w_2$ in this figure is expected to be much lower than $w_1$ and $w_3$. By doing so, this framework enables the policy network to efficiently extract information from reliable messages with reduced interference.}
  \label{fig:intro}
\end{figure*}
Adversarial attacks and defenses in communicative MARL receive much less attention when compared to their counterparts in reinforcement learning \cite{mu2022certified}. Current frameworks in this area \cite{ishii2022overview, yuan2023communication} commonly follow the principle of passive defense, where agents treat all received messages equally and try to make relative safe decisions. Since perturbed messages are mixed with useful messages, this indiscriminate reception may lead to the result of ``garbage in, garbage out''. 

We notice that robust communicative MARL has an important feature compared with robust RL: The attackers are only allowed to modify a part of the messages, while the modification is unlimited and perturbed messages can be quite different from the original ones. Inspired by noisy learning \cite{han2018co}, we propose an active defense strategy to utilize this feature: agents actively judge the reliability of messages based on their own unperturbed observations and hidden states (which contain history information) and reduce unreliable messages' impact on the final decision.\footnote{An intuitive approach is identifying the perturbed messages and simply remove them from decision-making. However, for an agent, the received messages themselves contain some unknown information, making it infeasible to accurately judge whether a message should be completely trusted or ignored. Besides, it is hard to determine a proper decision threshold for judging whether a message is reliable or unreliable.} For example, if in a search task an agent receives a message saying that ``target is at coordinates (1,1), get to it!'', while the agent have searched (1,1) and found nothing, it can realize that this message is fake. We also visualize the difference between active defense and passive defense in Fig.~\ref{fig:intro} for better demonstration.

Nevertheless, there remain two key challenges to implement active defense. (I) What kind of messages should be defined as unreliable? (II) How to adjust the unreliable messages' impact on the final decision properly? Since the ultimate goal for multi-agent communication is to make agents better finish cooperation tasks, a message beneficial for this goal should be considered reliable. However, common communicative MARL frameworks aggregate received messages with highly nonlinear neural networks, making it hard to evaluate or adjust the impact of a sole message on the final decision. 

To address the aforementioned challenges and achieve active defense, we introduce an Active Defense Multi-Agent Communication (ADMAC) framework  comprising two key components. 
The first is reliability estimator, which outputs the estimated reliability (ranging from 0 to 1) of a certain message according to the agent's observation and hidden state. The second is decomposable message aggregation policy net, which decomposes the impact of each message on the final decision, and allows adjustable weights to control this impact. Below is an introduction of how an agent processes messages and makes decisions using ADMAC: Firstly, it generates a \textbf{base action preference vector} based on the observation and hidden state, with each component corresponding to a feasible action. Secondly, for each message received, the agent generates a \textbf{message action preference vector} according to the message and the agent's observation. Thirdly, the reliability estimator outputs the reliability values of messages, which then serve as the weights of corresponding message action preference vectors. Lastly, the agent adds all weighted preference vectors to get the total action preference vector and feeds it into Softmax to derive the final action distribution. 

In order to substantiate the robustness of ADMAC, we evaluate it alongside three alternative communication frameworks in three communication-critical cooperation tasks. Besides, we implement four distinct types of attacks, including Gaussian attack \cite{he2023robust}, Monte-Carlo adversarial attack, Fast Gradient Sign Method \cite{goodfellow2014explaining} and Projected Gradient Descent \cite{madry2018towards}. Experiments confirm the superiority of ADMAC in terms of its robustness and performance compared to the alternatives. Moreover, we conduct ablation study to further explore the features of components in ADMAC. 

Our contributions can be summarized below:
\begin{enumerate}
    \item We elucidate a significant distinction between robust RL and robust communicative MARL, highlighting the unsuitability of the commonly employed passive defense strategy in robust RL for the latter modeling. Moreover, we propose the idea of active defense which exploits the features of robust communicative MARL.
    \item There remain two challenges to implement active defense, that are defining unreliable messages and adjusting the unreliable messages' impact on the final decision properly. To overcome them, we propose ADMAC, which incorporates mechanisms that enable agents to lower the impact of potentially malicious perturbed messages on the final decision, thereby achieving robustness against attacks.
    \item We empirically demonstrate the superiority of ADMAC over existing approaches, and conduct ablation study to delve deeper into the unique features and characteristics of it.
\end{enumerate}

\section{Preliminaries}
\subsection{Dec-POMDP with Communication}
Our modeling is based on a decentralized partially observable Markov decision process \cite{littman1994markov} with $N$ agents in the system. At timestep $t$, the global state is $s^t$, and agent $i$ receives observation $o_i^t$ from the environment and chooses an action $a^t_i$ to perform. Then the environment provides agents with rewards $r_i^t$ and updates the global state from $s^{t}$ to $s^{t+1}$ according to all agents' actions. 

There are numerous communication frameworks in communicative MARL, including scheduler-based \cite{rangwala2020learning}, inquiry-reply-based \cite{ma2021learning} and GNN-based \cite{pu2022attention} ones. They basically follow the communication and decision process specified below, regardless of the difference in communication architectures. 

At timestep $t$, agent $i$ with hidden states $h_i^{t-1}$ first receives $o_i^t$, based on which it generates message and communicate with other agents. Then, it updates its hidden states and chooses action $a^t_i$ with a policy network according to its current hidden state, observation and messages from others $m_1^t,m_2^t,...,m_N^t$. The objective function that needs to be maximized for agent $i$ is the discounted return:
\begin{equation}
\label{eqa:target}
    J(\theta) = \mathbb{E} [\sum_t \gamma^t r^t_i|\theta],
\end{equation}
where $\theta$ denotes the neural network parameters of the agents.

\subsection{Attack against Communication}
Attack against multi-agent communication has been a hot topic \cite{xue2022mis,sun2023certifiably} recently since wireless communication is vulnerable to distractions and noises. These works commonly follow the setting that attackers only interfere a part of messages in the multi-agent system to disrupt collaboration. Based on them, we consider the following attack model.

Suppose there are $N$ agents in the system, and at each timestep there are at most $N\times N$ messages. For each message $m_j^t$, the attacker has a probability $p$ to change it to $\hat{m}_j^t$. Then agents receive messages without knowing which are perturbed. 

The attack may be adversarial or non-adversarial, depending on how much information the attacker has about the agents. For adversarial attack, we adopt the setting of \citeauthor{zhang2020robust} where the attacker tries to minimize the chosen probability or preference of the best action, denoted by $\hat{P}(a^t_{i,best})$:
\begin{equation}
\label{eqa:minbestp}
\hat{m}_j^t = f_A(m_j^t) = \arg\min \sum_{i\neq j} \hat{P}(a^t_{i,best}).
\end{equation}
Another feasible attack target is to maximize the KL divergence between $a^t_i$ and $\hat{a}^t_i$, with $a^t_i$ representing the action distribution output by agent $i$ when receiving raw message $m_j^t$ and $\hat{a}^t_i$ representing the action distribution output by agent $i$ when receiving perturbed message $\hat{m}_j^t$:
\begin{equation}
\label{eqa:maxkl}
\hat{m}_j^t = f_B(m_j^t) = \arg\max \sum_{i\neq j} D_{KL}(\hat{a}^t_i|a^t_i).
\end{equation}

Following the setting of \cite{sun2023certifiably}, we consider a strong attack model to better evaluate the robustness of models: The strength of the perturbation, denoted as $||\hat{m}_j^t-m_j^t||$, is not bounded. Consequently, there are multiple ways to implement $f_A(\cdot)$ and $f_B(\cdot)$, which will be detailed in the Experiments Section.


\section{Active Defense Multi-Agent Communication Framework}
\begin{figure*}[t!]
  \centering
  \includegraphics[width=0.95\textwidth]{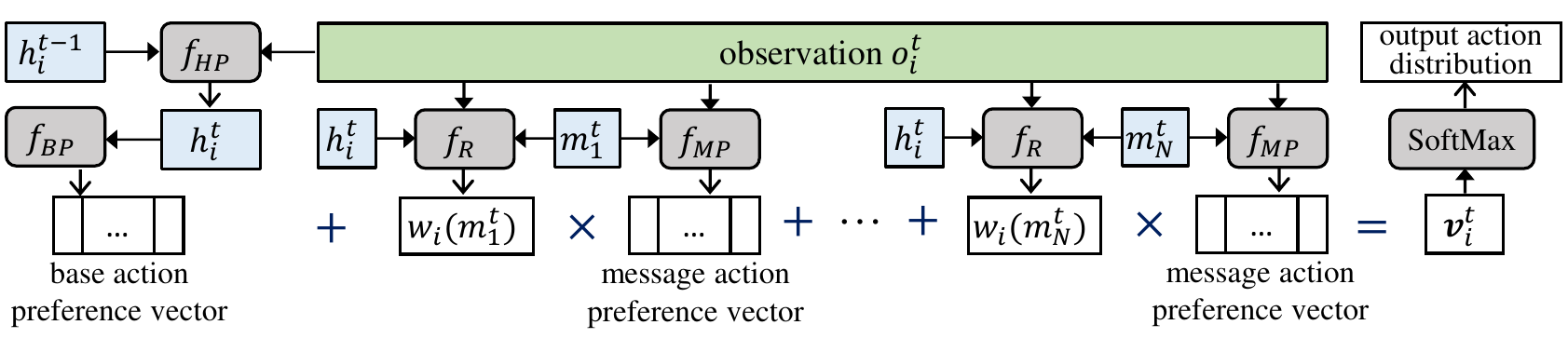}
  \caption{The figure shows how agent $i$ with hidden state $h_i^{t-1}$ generates an output action distribution within ADMAC after receiving an observation $o_i^t$ and messages $m_1^t$, $m_2^t$, ..., $m_N^t$ from others. It is noteworthy that the length of actions preference vectors is the same as the number of feasible actions, and each component of the vectors represents the preference for the corresponding action. As depicted in the figure, the impact of each message on the final decision is restricted to the respective action preference vector and can be regulated by the weight $w_i(m_j^t)$.}
  \label{fig:alg}
\end{figure*}

We propose an active defense idea for robustness, namely, making agents judge the reliability of received messages based on their local information and reduce unreliable messages' impact on the final decision. However, the implementation brings two challenges: (I) How to define ``unreliable'' messages? (II) How to adjust the unreliable messages' impact on the final decision? In this section, we propose ADMAC comprising a decomposable message aggregation policy net and a reliability estimator to address these two challenges, whose visualization is presented in Fig.\ref{fig:alg}.

\subsection{Decomposable Message Aggregation Policy Net}
The decomposable message aggregation policy net $f_P$ is designed to decompose the impact of each message on the final decision by restricting their influence to action preference vectors. Specifically, $f_P$ consists of three parameterized modules: a GRU module $f_{HP}$ used to update hidden states with observations, a base action generation module $f_{BP}$ that generates \textbf{base action preference vectors} according to the updated hidden states, and a message-observation process module $f_{MP}$ that generates \textbf{message action preference vectors} according to the observations and received messages. Suppose there are $K$ actions to choose from, then an action preference vector has $K$ components, each representing the preference for the corresponding action. Use $p^t_i = f_P(o_i^t, h_i^t, m_1^t,..., m_N^t)$ to denote the final output action distribution for agent $i$ at timestep $t$, where the $k$-th component of $p^t_i$, denoted by $p^t_i[k]$, refers to the probability of choosing the $k$-th action, the decision process is formulated below:
\begin{equation}
    \label{eqa:decision}
        \begin{split}
            h_i^t &= f_{HP}(h_i^{t-1},o_i^t), \\
            v^t_i &= f_{BP}(h_i^t) + \sum_{j\neq i} 
     w_i(m_j^t)f_{MP}(o_i^t, m_j^t), \\
            p^t_i[k] &= e^{v^t_i[k]}/\sum_k e^{v^t_i[k]}.
        \end{split}
    \end{equation}
where $v^t_i$ is the \textbf{total action preference vector}, and $w_i(m_j^t)$ is a weight determined by the reliability estimator detailed in the next subsection. $w_i(m_j^t)$ is set to $1$ by default if no robustness is required. Evidently, messages with larger weights have a stronger influence on the final decision. Therefore, for a message considered to be malicious, reducing its weight can effectively attenuate its impact on the final decision. We provide the following proposition to characterize this feature: 
\begin{proposition}
\label{prop1}
  For an agent making decisions using (\ref{eqa:decision}), if message $m_j^t$ recommends an action most or least recommends an action to the agent, incorporating this message into decision-making must increase or decrease the probability of choosing this action in the final decision, and the magnitude of this effect varies monotonically with the weight $w_i(m_j^t)$.
\end{proposition}
The proof is presented in Appendix A. Here ``$m_j^t$ recommends the $k_a$-th action most'' means $k_a=\arg\max_k f_{MP}(o_i^t, m_j^t)[k]$, and ``$m_j^t$ recommends the $k_b$-th action least'' means $k_b=\arg\min_k f_{MP}(o_i^t, m_j^t)[k]$.

\subsection{Reliability Estimator}
The reliability estimator is a classifier $f_R(h_i^t, o_i^t, m_j^t)$ that judges whether a message $m_j^t$ is reliable for agent $i$ with the help of the agent's hidden state $h_i^t$ and observation $o_i^t$. The output of $f_R(h_i^t, o_i^t, m_j^t)$ is a vector of length $2$ normalized by Softmax. Let
\begin{equation}
    w_i(m_j^t) = f_R(h_i^t, o_i^t, m_j^t)[0].
\end{equation}
$w_i(m_j^t)$ represents the extent to which the reliability estimator thinks $m_j^t$ is reliable for agent $i$, and is used as the weight of $m_j^t$ in (\ref{eqa:decision}). Besides, judging the reliability of messages can be treated as a binary classification problem, and the key challenge here is labelling data, i.e., defining what messages are ``reliable'' and what are not. Since the ultimate goal of communicative MARL is maximizing cooperation performance represented by (\ref{eqa:target}), messages should be labelled according to whether they are conducive to achieving this goal. 

With the assumption that agents' policies are well-trained, we use the following criteria to label messages: For an agent, if a received message recommends it to choose the best action, then the message is considered to be reliable, otherwise it is bad. Here ``best action'' refers to the action most likely to be chosen in the absence of perturbations, and `` $m_j^t$ recommends the $k_r$-th action'' is defined as $f_{MP}(o_i^t, m_j^t)[k_r]>\sum_k f_{MP}(o_i^t, m_j^t)[k]/K $. With the labelled messages, the reliability estimator can be trained in a supervised learning way.

\subsection{Instantiation}
ADMAC only specifies how agents process messages as well as observations to make robust decisions, and is compatible with numerous communication architectures and training algorithms. In this paper, we adopt a basic broadcast communication mechanism \cite{singh2018learning} and use the following paradigm to train our model. More details are provided in Appendix B. 

\textbf{Stage 1: training policy net}: The training goal of this stage is to optimize the parameters of the decomposable message aggregation policy net and message encoders to maximize the objective function defined in (\ref{eqa:target}). Notably, messages used in our framework are one-dimensional vectors with each component ranging from $-1$ to $1$, and this stage does not involve any attacks or defenses. Consequently, the agents interact with the environment, communicate with each other and update the parameters in a traditional communicative MARL way.  

\textbf{Stage 2: generating dataset}: After the policies are well-trained, let agents interact with the environment for several episodes to generate training dataset for reliability estimator. To enhance its identification ability, we implement two representational and strong attacks from \cite{sun2023certifiably} during training:  

\textbf{Attack I. Random Perturbation}: It suits the scenario where the attacker has no information about the agents. Attackers generate random perturbed messages $\hat{m}_j^t$ to replace the original ones, which means each component of $\hat{m}_j^t$ is a sample from Uniform distribution on $(-1,1)$. 

\textbf{Attack II. (L2-normed) Gradient descent adversarial attack}: It suits the scenario where the attacker has full knowledge of the agents. To maximize the adversarial attack objective presented in (\ref{eqa:minbestp}), gradient descent can be utilized to generate a perturbed message:
\begin{equation}
    \hat{m}_j^t = m_j^t + \lambda \nabla_{m_j^t} f_A(m_j^t)/||\nabla_{m_j^t} f_A(m_j^t)||_2
\end{equation}

When creating dataset for the reliability estimator, we randomly replace 1/3 raw messages with Attack I messages and 1/3 messages with Attack II messages, and let agents make decisions based on them. The decisions will not be truly executed, on the contrary, they are only used to label these messages according to the aforementioned criteria (It is possible that some perturbed messages are labelled reliable, and unperturbed messages are labelled unreliable). All messages and corresponding observations, hidden states, and labels are collected to form a dataset for supervised learning.

\textbf{Stage 3: training reliability estimator}: Train the reliability estimator parameterized via MLP based on the dataset using Adam optimizer and cross-entropy loss function. A well-trained reliability estimator is expected to assign low weights to messages that are not conducive to an agent's selection of the optimal action.

\begin{figure*}[t!]
    \centering
    \begin{minipage}[t]{0.32\textwidth}
    \centering
    \includegraphics[width=1\textwidth]{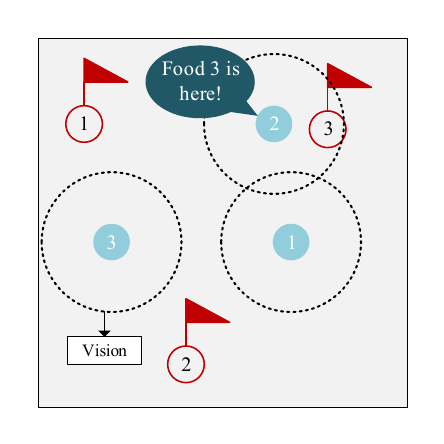}
    \end{minipage}
    \begin{minipage}[t]{0.32\textwidth}
    \centering
    \includegraphics[width=1\textwidth]{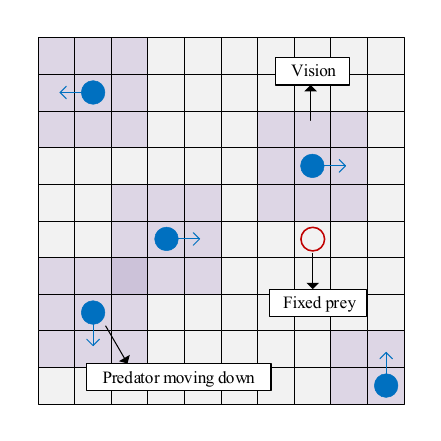}
    \end{minipage}
    \begin{minipage}[t]{0.32\textwidth}
    \centering
    \includegraphics[width=1\textwidth]{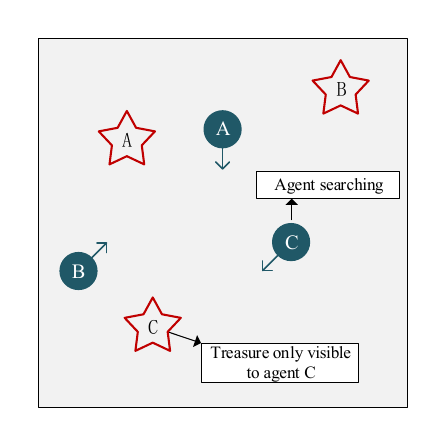}
    \end{minipage}
    \caption{Visualizations of our three experiment environments: Food Collector, Predator Prey, and Treasure Hunt.}
    \label{fig:env}
\end{figure*}

\section{Experiments}
\subsection{Experimental Environments}
We implement three communication-critical multi-agent environments for demonstrative purposes: Food Collector \cite{sun2023certifiably}, Predator Prey \cite{singh2018learning}, and Treasure Hunt \cite{freed2020communication}. They either adhere to a predefined communication setting or a learned communication setting. We test these two kinds of communication settings because the messages within them have different distributions, potentially influencing the performance of attack and defense. Within all environments, one episode ends if all agents finish their tasks or the timestep reaches the upper limit $t_{max}$. Therefore, lower average timesteps indicate better performance. These environments are detailed below and visualized in Fig.\ref{fig:env}.

\textbf{Food Collector (predefined communication)} In this task, $N=5$ agents with different IDs search for foods with the same IDs in a $1\times1$ field.  Each agent can observe targets within its vision $d=0.2$, and moves in eight directions at a speed $v=0.15$. If an agent finds a target with a different ID, it will kindly broadcasts the coordinates to help the corresponding agent find it. 

\textbf{Predator Prey (learned communication)} In this task, $N=5$ agents with vision $1$ are required to reach a fixed prey in a grid world of size $10 \times 10$. Due to the severely limited perception, agents must communicate with others to finish tasks earlier. For example, they can delineate their respective search areas through communication, and the first agent to reach the prey can tell others the coordinates.

\textbf{Treasure Hunt (learned communication)} In this task, $N=5$ agents work together to hunt treasures in a field of size $1 \times 1$ with movement speed $v=0.9$. Each agent obtains the coordinates of its own treasure, which is invisible to others. Note that an agent cannot collect its treasure by itself. Instead, it should help others hunt it through learned communication.

\subsection{Tested Algorithms}
We evaluate our proposed ADMAC alongside three alternatives introduced below:

\textbf{Targeted Multi-Agent Communication (TARMAC)} \cite{das2019tarmac} It is a baseline communication framework where agents utilize an attention mechanism to determine the weights of receiving messages. It does not include any robust techniques, and its performance demonstrates the impact of attacks without any defense.

\textbf{Adversarial Training (AT)} \cite{pattanaik2018robust,tu2021adversarial} This framework adopts the most frequently used robust learning technique, which is performing adversarial attacks during training to gain robustness.

\textbf{Ablated Message Ensemble (AME)} \cite{sun2023certifiably} It constructs a message-ensemble policy that aggregates multiple randomly ablated message sets. The key idea is that since only a small portion ($<50\%$) of received messages are harmful, making agents take the consensus of received messages should provide robustness.

All tested frameworks are trained with an improved version of REINFORCE \cite{williams1992simple}. With the assumption that all agents of one task are homogeneous, they share policy network and reliability estimator parameters to accelerate training. More information about training is presented in Appendix B. 

\subsection{Implemented Attacks}
When conducting experimental evaluations of the models, we implement the following four kinds of attacks, including adversarial and non-adversarial ones:

\textbf{Attack III Gaussian attack} \cite{he2023robust}: Attacker adds Gaussian Noises to the messages, i.e. 
\begin{equation}
    \hat{m}_j^t = m_j^t + \sigma N(0,1).
\end{equation}
We set $\sigma=0.5$ in the experiments. 

\textbf{Attack IV Monte-Carlo adversarial attack}: Randomly generate 10 messages, and find the one that maximizes the attack objective $f(m_j^t)$ defined in (\ref{eqa:minbestp}) or (\ref{eqa:maxkl}). 

\textbf{Attack V Fast Gradient Sign Method}\cite{goodfellow2014explaining}: 
\begin{equation}
    \hat{m}_j^t = m_j^t + \eta sign(\nabla_{m_j^t} f(m_j^t)), 
\end{equation}
Then range of $m$ is $(-1,1)$, and we set $\eta=1$ to obtain a strong attack.

\textbf{Attack VI Projected Gradient Descent} \cite{madry2018towards}: PGD can be treated as a multi-step version of FGSM:
\begin{equation}
    \hat{m}_j^{t,(i+1)} = \hat{m}_j^{t,(i)} + \epsilon sign(\nabla_{\hat{m}_j^{t,(i)}} f(\hat{m}_j^{t,(i)})), 
\end{equation}
where $\hat{m}_j^{t,(0)}=m_j^t$. We set $\epsilon=0.3$ and use 5-step updates to obtain the final perturbed messages.

For adversarial attacks (IV, V and VI), the attack objective can be chosen from  $f_A(\cdot)$ and $f_B(\cdot)$, defined accordingly in (\ref{eqa:minbestp}) and (\ref{eqa:maxkl}). It is noteworthy that although they may have the same objectives with attack-II, which is used to train ADMAC, the generated perturbed messages belong to different distributions because they are calculated differently. In other words, our ADMAC framework DOES NOT have prior information of the above attacks used for evaluation.

\begin{figure*}[t!]
    \centering
    \begin{minipage}[t]{0.24\textwidth}
    \centering
    \includegraphics[width=1\textwidth]{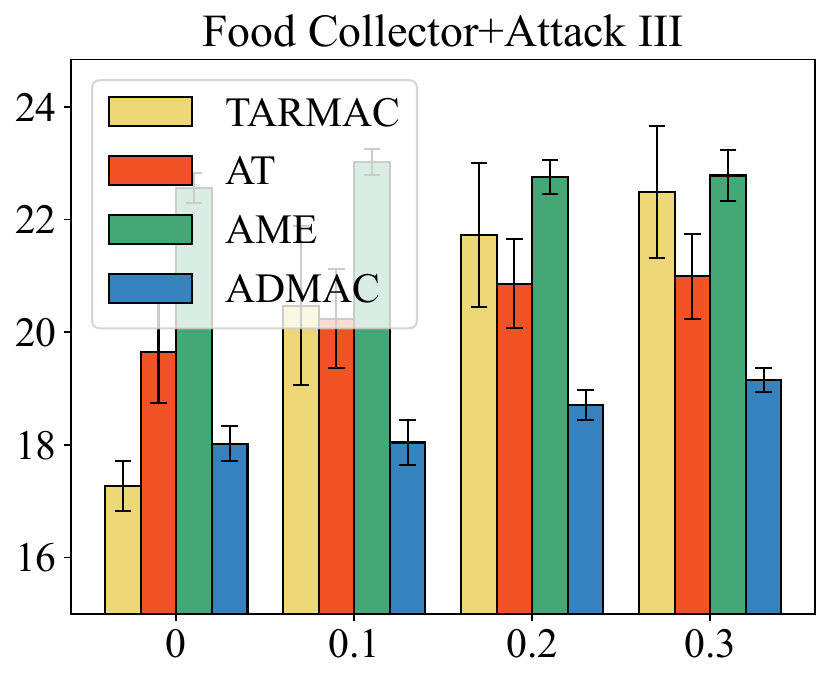}
    \end{minipage}
    \begin{minipage}[t]{0.24\textwidth}
    \centering
    \includegraphics[width=1\textwidth]{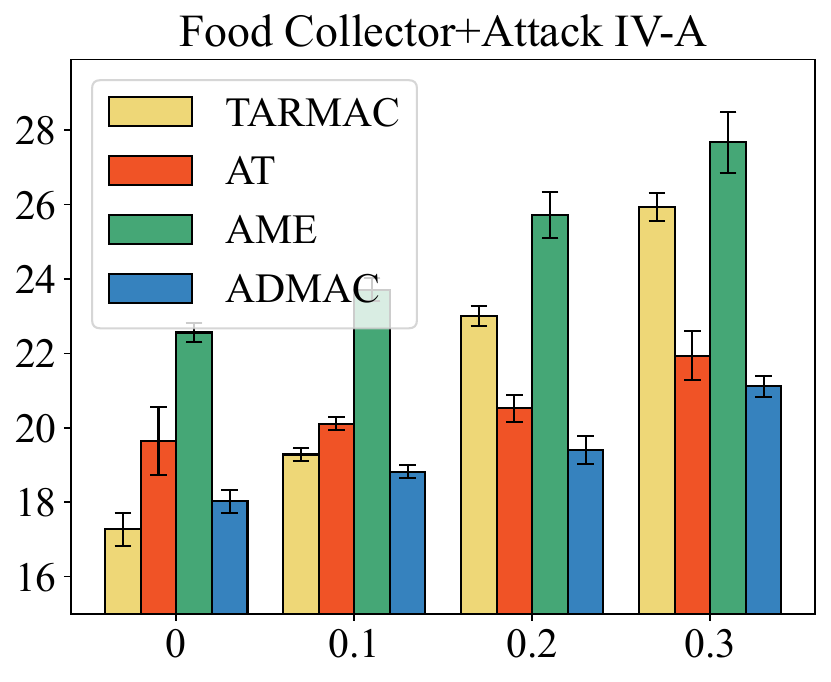}
    \end{minipage}
    \begin{minipage}[t]{0.24\textwidth}
    \centering
    \includegraphics[width=1\textwidth]{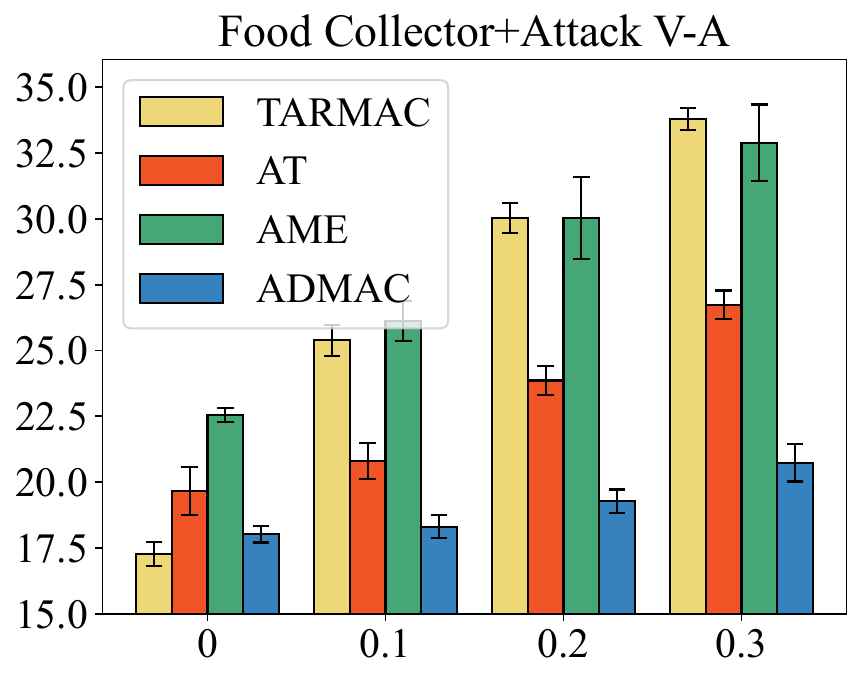}
    \end{minipage}
    \begin{minipage}[t]{0.24\textwidth}
    \centering
    \includegraphics[width=1\textwidth]{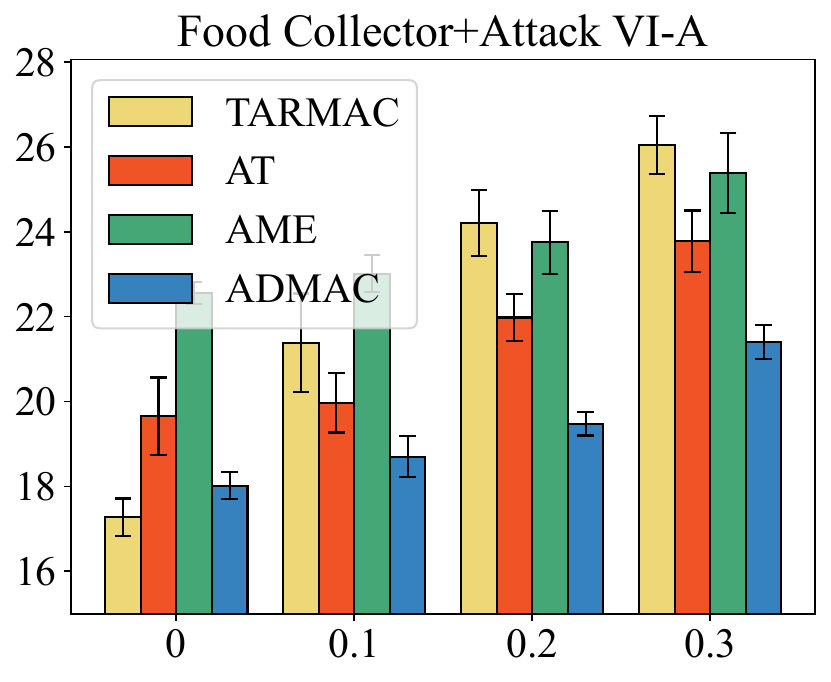}
    \end{minipage}    
    \begin{minipage}[t]{0.24\textwidth}
    \centering
    \includegraphics[width=1\textwidth]{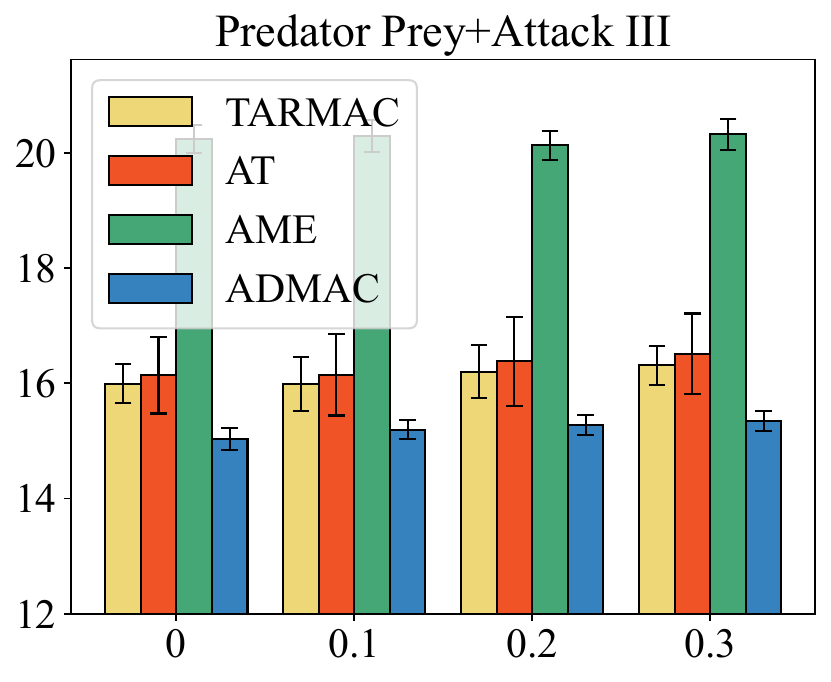}
    \end{minipage}
    \begin{minipage}[t]{0.24\textwidth}
    \centering
    \includegraphics[width=1\textwidth]{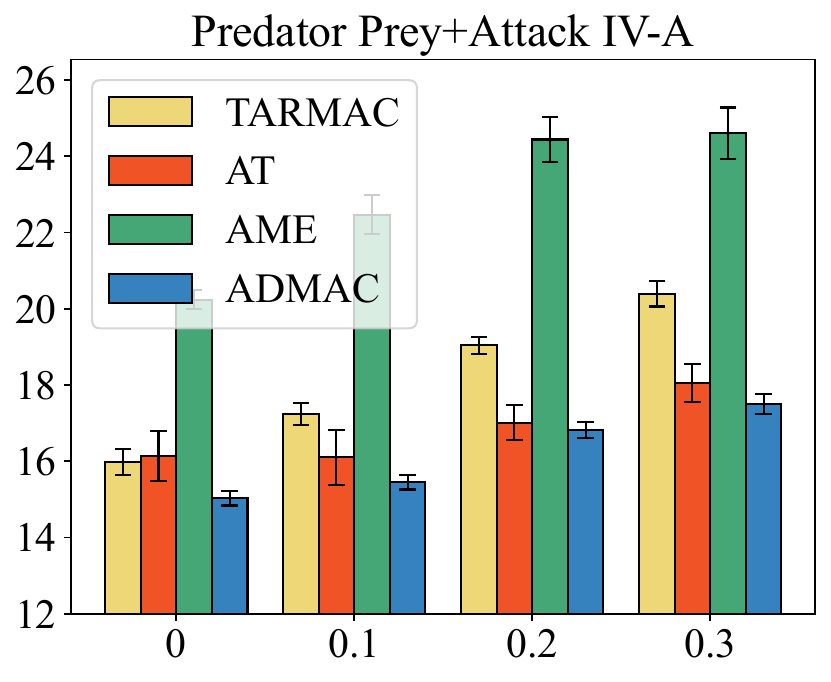}
    \end{minipage}
    \begin{minipage}[t]{0.24\textwidth}
    \centering
    \includegraphics[width=1\textwidth]{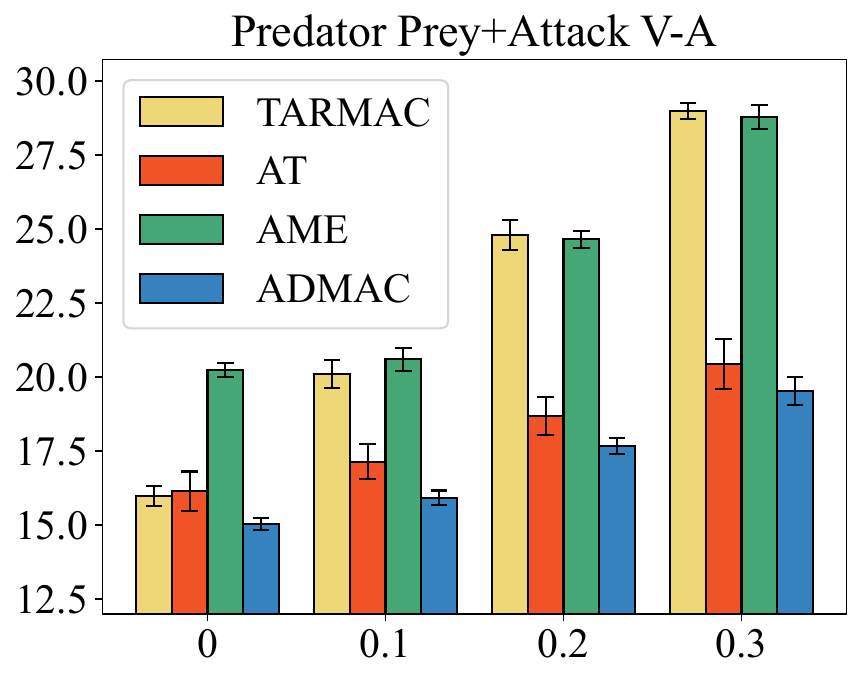}
    \end{minipage}
    \begin{minipage}[t]{0.24\textwidth}
    \centering
    \includegraphics[width=1\textwidth]{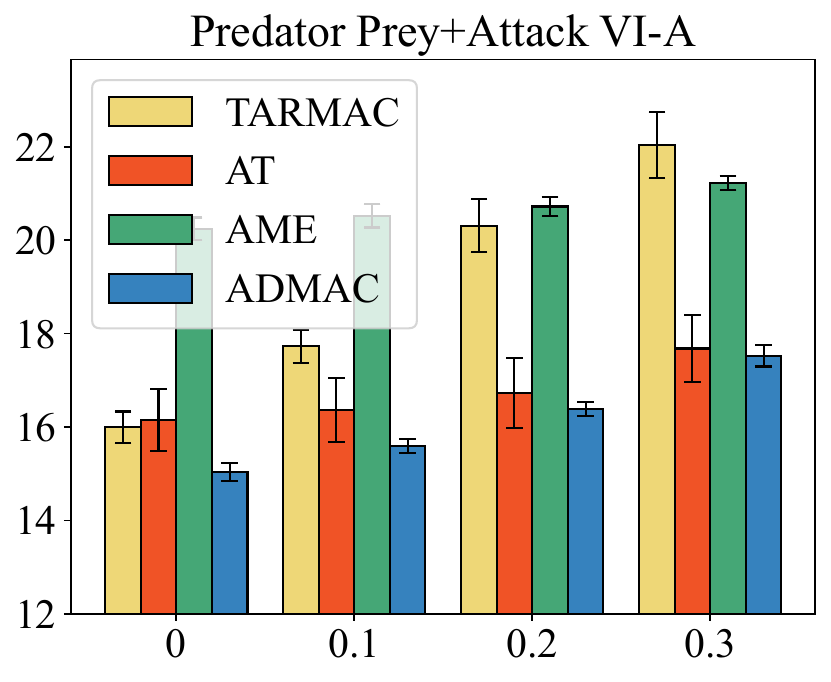}
    \end{minipage}   
    \begin{minipage}[t]{0.24\textwidth}
    \centering
    \includegraphics[width=1\textwidth]{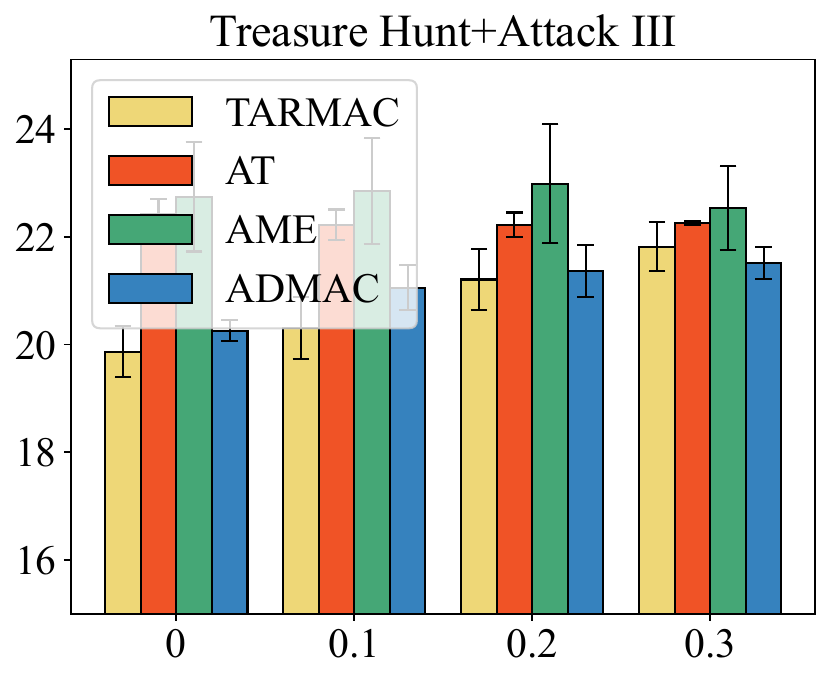}
    \end{minipage}
    \begin{minipage}[t]{0.24\textwidth}
    \centering
    \includegraphics[width=1\textwidth]{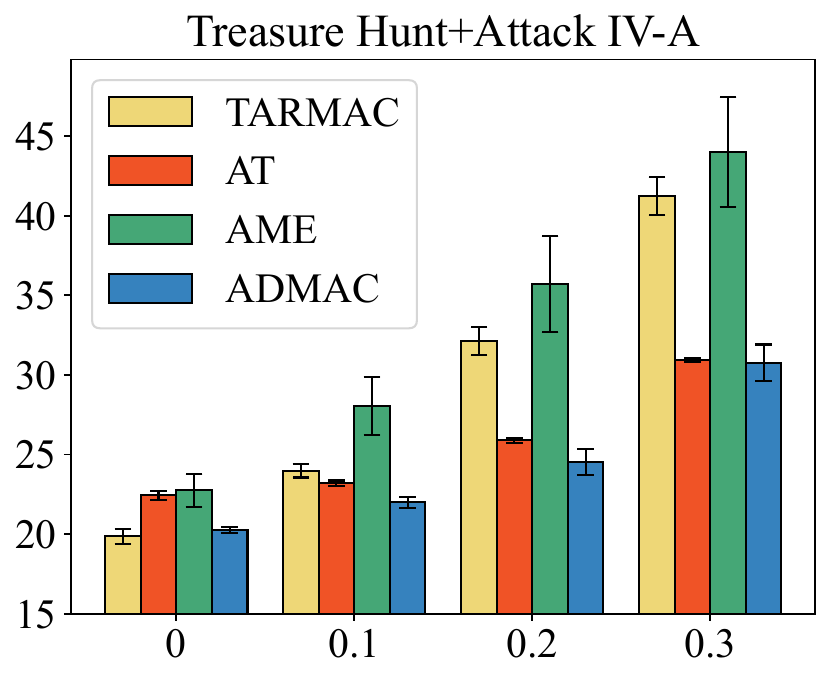}
    \end{minipage}
    \begin{minipage}[t]{0.24\textwidth}
    \centering
    \includegraphics[width=1\textwidth]{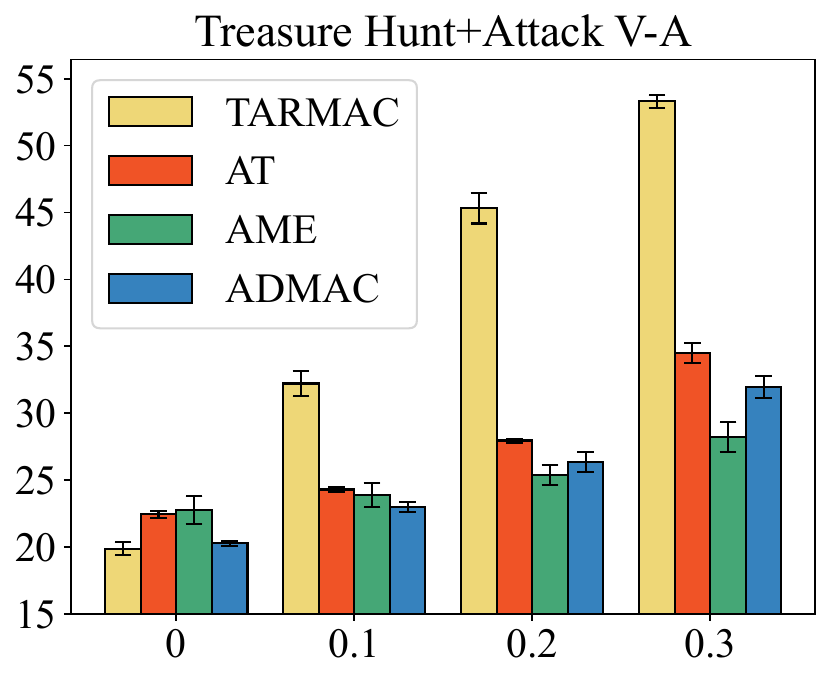}
    \end{minipage}
    \begin{minipage}[t]{0.24\textwidth}
    \centering
    \includegraphics[width=1\textwidth]{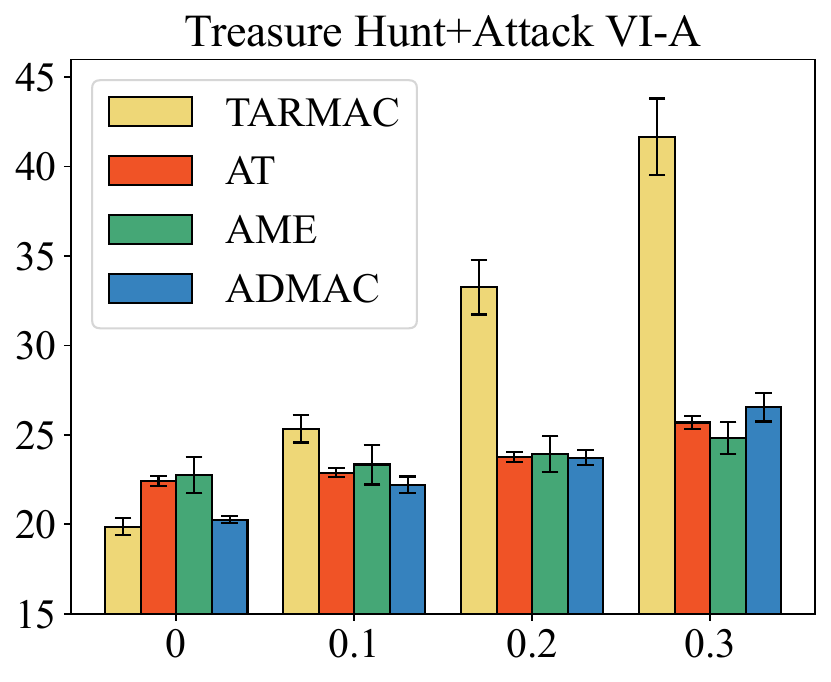}
    \end{minipage}       
    \caption{Results for the main experiments. The x-axis represents the attack probability, and the y-axis represents the timesteps required for task completion, where smaller values signify better performance. For each setting (e.g. TARMAC in Treasure Hunt), we train five models with different seeds. During test, we run 500 episodes for each setting and plot the mean as well as standard error of the averaged timesteps. Lower timesteps indicate better performance.}
    \label{fig:mainexp}
\end{figure*}
\subsection{Performance under Attacks}
We evaluate the four multi-agent communication frameworks with the aforementioned tasks and attacks. Additionally, we set the attack probability $p$ to different values to observe how the performance of the agents decays as the attack intensity increases. Besides, since attack objective $f_B(\cdot)$ is not applicable to AME, we only present adversarial attacks with objective function $f_A(\cdot)$ in the main experiments, and provide additional experimental results in Appendix C. From the results presented in Fig.\ref{fig:mainexp}, the following conclusions can be drawn. 

Attack III is the weakest attack, and its attack effect on TARMAC (the baseline non-robust framework) indicates the extent to which messages can influence decision-making, which also reflects the importance of communication for cooperation in specific attacks. It can be seen that communication is important in all tasks, with the order of importance being Treasure Hunt $>$ Food Collector $>$ Predator Prey. 

AT is one of the most popular techniques in robust learning, and it is empirically confirmed to be a relatively reliable method. Compared with TARMAC, AT achieves lower timesteps under strong attacks, however, its baseline performance without attacks is slightly worse. This is inevitable as the trade-off between robustness and performance has long been a concern in adversarial training. 

Though AME provides robustness against attacks to a certain degree, its baseline performance without attacks is bad, severely dragging down the overall performance. This results from the key feature of AME: adopting the consensus of all incoming messages. It protects the agents from being influenced by a small number of malicious messages, but it also prevents agents from getting exclusive information. 

Our proposed framework ADMAC has the best overall performance, and this advantage is provided by the active defense strategy. Compared with AT and AME, ADMAC exploits the features of adversarial multi-agent communication, judging received messages' reliability based on agents' observations and hidden states. Notably, ADMAC has better baseline performance than TARMAC in Predator Prey tasks, which is caused by the feature of decomposable message aggregation policy and will be discussed more in the ablation study. 

\begin{figure*}[t!]
    \centering
    \begin{minipage}[t]{0.32\textwidth}
    \centering
    \includegraphics[width=1\textwidth]{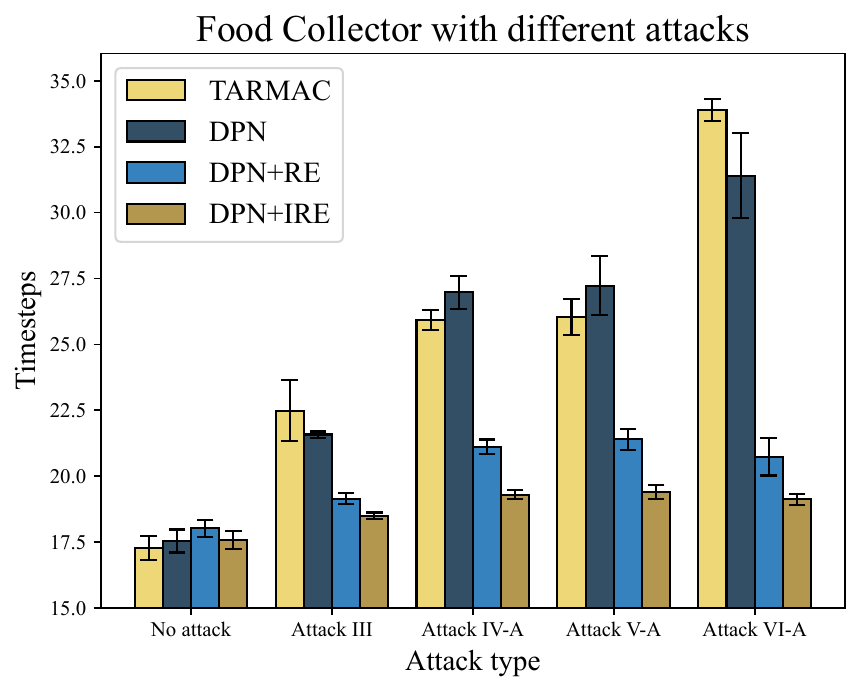}
    \centerline{RE recall/precision: 0.90/0.86}
    \end{minipage}
    \begin{minipage}[t]{0.32\textwidth}
    \centering
    \includegraphics[width=1\textwidth]{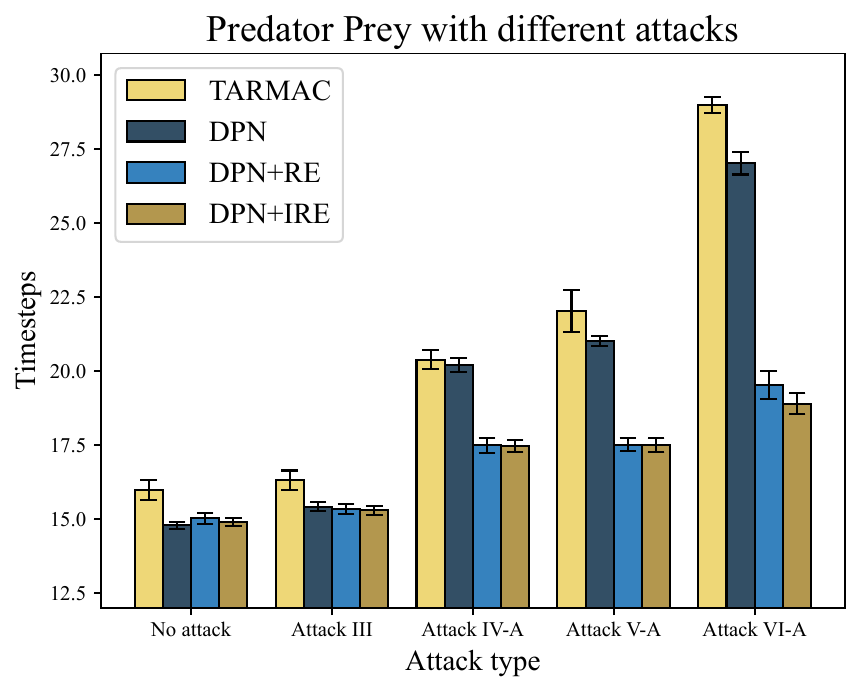}
    \centerline{RE recall/precision: 0.87/0.84}
    \end{minipage}
    \begin{minipage}[t]{0.32\textwidth}
    \centering
    \includegraphics[width=1\textwidth]{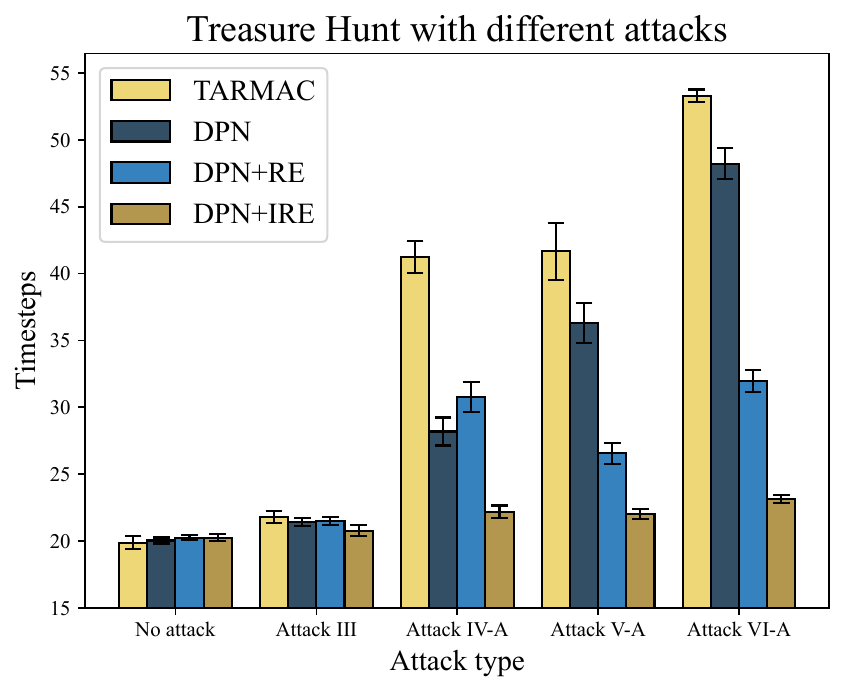}
    \centerline{RE recall/precision: 0.79/0.76}
    \end{minipage}

    \caption{Results for the ablation study. The attack probability is set to $p=0.3$ for all types of attack.}
    \label{fig:ablation1}
\end{figure*}

\subsection{Ablation Study}
ADMAC consists of two components: the reliability estimator (abbreviated as RE) and the decomposable message aggregation policy net (abbreviated as DPN). To further investigate each component's function, we evaluate the following four methods under attacks: TARMAC, DPN only, DPN+RE (i.e. ADMAC), and DPN+IRE. Here IRE refers to an ideal reliability estimator, achieving $100\%$ classification accuracy and cannot be implemented in reality. We also list the recall and precision of RE below the figures to analysis its performance from the perspective of supervised learning. 

From the results presented in Fig~\ref{fig:ablation1}, the following conclusions can be drawn. Firstly, the special structure of DPN does not degrades baseline performance, on the contrary, it provides a little robustness sometimes. Secondly, RE provides considerable robustness, especially against adversarial attacks. Thirdly, DPN+IRE has the best robust performance, which again verifies the effectiveness of the way we label messages. Additionally, the recall and precision of RE accounts for the performance gap between DPN+RE and DPN+IRE: The better the classification performance of RE is, the closer the performance of DPN+RE is to DPN+IRE. 


\section{Related Work}

TMC \cite{zhang2020succinct} is one of the earliest robust multi-agent communication frameworks. However, it only provides certain robustness against Gaussian noises and random message loss. Some recent researches have considered the existence of malicious attackers \cite{blumenkamp2021emergence, tu2021adversarial, yuan2023communication, sun2023certifiably}, who perturb normal messages or send fake messages to disrupt normal operation of multi-agent systems. Agents within these frameworks commonly receive all messages equally and try to make robust decisions, which follows a passive defense idea conventional in robust RL \cite{havens2018online, pattanaik2018robust}. Notably, passive defense strategies fail to utilize one important feature of robust communicative MARL: unperturbed messages and hidden states can help identify fake messages to some extent. Unlike the aforementioned methods, R-MACRL \cite{xue2022mis} takes a rather proactive defense strategy, which is correcting perturbed messages. Since useful messages must contain some information unknown to the agents and attackers may replace the useful messages with manipulated ones, this strategy is not quite practical. Compared with them, while utilizing observations and hidden states, we have made corresponding preparations for the fact that it is impossible to perfectly judge received messages.

Besides, we would like to differentiate our research from robust MARL. Following \citeauthor{sun2023certifiably}, we make two important assumptions: 1) only a part of the messages might be perturbed; 2) the perturbation power is unlimited and the perturbed messages can be completely different from the original ones. As a comparison, robust MARL commonly assumes the observations of agents \cite{li2019robust} or the environment model \cite{zhang2020robust2} is perturbed. Some robust MARL techniques can be used in robust communicative MARL (e.g. adversarial training), but an active defense strategy can better utilize the features of this problem.

\section{Conclusion}
In this paper, we investigate the issue of robust communicative MARL, where malicious attackers may interrupt the multi-agent communication, and agents are required to make resilient decisions based on unperturbed observations and potentially perturbed messages. Considering that agents have multiple sources of information in this setting, we put forward an active defense strategy, which involves agents actively assessing the reliability of incoming messages and reducing the impact of potentially malicious messages to the final decisions. 

Implementing active defense poses two primary challenges: labelling ``unreliable'' messages and adjusting the unreliable messages' impact on the final decision appropriately. To address them, we introduce ADMAC, a framework comprising an reliability estimator and a decomposable message aggregation policy net. ADMAC is evaluated alongside three alternative multi-agent communication frameworks in three communication-critical environments under attacks of different kinds and intensities, and shows outstanding robustness. 

One limitation of ADMAC is that in scenarios where messages are likely to carry unique information, judging whether a message is reliable is hard, leading to a decline in the robustness of ADMAC. As future work, we will try to make agents aggregate information from a wider range of sources in order to better assess received messages, and expand our framework to accommodate scenarios with continuous action spaces.




\appendix

\onecolumn
\section{A. Proof of Proposition 1}
Proposition 1 can be formulated as follows:

If
\begin{equation}
    \begin{split}
    \label{condition}
        v_i &= f_{BP}(h_i) + \sum_{j\neq i, u} w_i(m_j)f_{MP}(o_i, m_j), \ \ p_i[k] = e^{v_i[k]}/\sum_k e^{v_i[k]}, \\
        \hat{v}_i &= v_i + wf_{MP}(o_i, m_u), \ \ \hat{p}_i[k] = e^{\hat{v}_i[k]}/\sum_k e^{\hat{v}_i[k]}, \\
        k_{max} &= \arg\max_k f_{MP}(o_i, m_u)[k], \\ 
        k_{min} &= \arg\min_k f_{MP}(o_i, m_u)[k], \\ 
    \end{split}
\end{equation}
then
\begin{equation}
        \partial \hat{p}_i[k_{max}] / \partial w >0, \ \ \partial \hat{p}_i[k_{min}] / \partial w < 0    \\
\end{equation}

\textit{Proof}:

$f_{MP}(o_i, m_u)$ and $v_i$ both are vectors containing $K$ components, and we denote them as:
\begin{equation}
\begin{split}
    v_i &= [a_1,a_2,...,a_K] \\
    f_{MP}(o_i, m_j) &= [b_1,b_2,...,b_K] 
\end{split}
\end{equation}
Then $\hat{p}_i[k_{max}]$ is calculated below:
\begin{equation}
\begin{split}
    \hat{p}_i[k_{max}] &= \frac{e^{a_{k_{max}}+ w b_{k_{max}}}  }{ \sum_{k=1}^K e^{a_{k}+ w b_{k}}} \\
    &= \frac{1}{\sum_{k=1}^K e^{w (b_{k}- b_{k_{max}})} e^{a_{k}-a_{k_{max}}}} \\
    &= \frac{1}{G_{max}(w)}
\end{split}
\end{equation}
Notably,
\begin{equation}
    \partial G_{max}(w)/\partial w= \sum_{k=1}^K (b_{k}- b_{k_{max}})e^{w (b_{k}- b_{k_{max}})} e^{a_{k}-a_{k_{max}}}
\end{equation}
According to (\ref{condition}), $b_{k}- b_{k_{max}} \leq 0$. Besides, $e^{w (b_{k}- b_{k_{max}})} e^{a_{k}-a_{k_{max}}} >0$. Therefore, $\partial G_{max}(w)/\partial w<0$, and we get $\partial \hat{p}_i[k_{max}] / \partial w >0$. Similarly, $b_{k}- b_{k_{min}} \geq 0$ and $\partial \hat{p}_i[k_{min}] / \partial w < 0$.

\section{B. Implementation Details}
\subsection{Training details}
Our ADMAC framework consists of two key components, namely, a decomposable message aggregation policy net $f_P$ and a reliability estimator $f_R$. $f_P$ is trained using policy gradient methods specified below.

Use $\theta$ to denote the parameters of $f_P$, and the training goal is to maximize $J(\theta) = \mathbb{E} [\sum_t \gamma^t r^t_i|\theta]$. Use $\pi_{\theta}(h,o,m,a)$ to represent the probability of choosing action $a$ with hidden state $h$, observation $o$ and message $m$, $\nabla_{\theta}J(\theta)$ can be calculated using $N_e$ complete episodes consisting of $N_t$ transitions:
\begin{equation}
\label{eqa:gradient}
    \nabla_{\theta}J(\theta) = \sum_{j=1}^{N_t}\pi_{\theta}(h_j,o_j,m_j,a_j)A(h_j,o_j,m_j,a_j)\nabla_{\theta}\log\pi_{\theta}(h_j,o_j,m_j,a_j),
\end{equation}
where $h_j,o_j,m_j,a_j$ are from the $j$-th transition used for training and $A(h_j,o_j,m_j,a_j)$ is the advantage function calculated as follows:
\begin{equation}
    A(h_j,o_j,m_j,a_j) = Q(h_j,o_j,m_j,a_j) - V_{\phi}(h_j,o_j,m_j),
\end{equation}
where $Q(h_j,o_j,m_j,a_j)$ can be calculated using complete episodes in the replay buffer and $V_{\phi}(h_j,o_j,m_j)$ is approximated via neural networks with parameter $\phi$. Suppose the $j$-th transition is the $t_j$-th timestep of the episode it belongs to, then
\begin{equation}
\label{eqa:qcompute}
    Q(h_j,o_j,m_j,a_j) = \sum_{t=t_j}^{t_{max}} \gamma^{t-t_j} r^t.
\end{equation}
$\phi$ is optimized by minimizing the value loss:
\begin{equation}
\label{eqa:value_loss}
    L_V(\phi) = \sum_{j=1}^{N_t}(Q(h_j,o_j,m_j,a_j)-V(h_j,o_j,m_j))^2.
\end{equation}

$f_R$ with parameter $\psi$ is trained in a typical supervised learning way. Suppose there are $N_s$ samples in the dataset, and the $i$-th sample comprises $x_i = [h_i,o_i,m_i]$ and $y_i = 0$ or $1$, where $y_i=0$ represents the message is considered reliable and $y_i=1$ represents the opposite. Then the loss is calculated as follows:
\begin{equation}
\label{eqa:sl}
    L_R(\psi) = -\sum_{i=1}^{N_s} (y_i \log f_R(h_i,o_i,m_i)[1] + (1-y_i) \log f_R(h_i,o_i,m_i)[0]).
\end{equation}
We present the pseudo code for our framework in Algorithm~\ref{alg: ours}. 

\begin{algorithm}[t!]
	\renewcommand{\algorithmicrequire}{\textbf{Input:}}
	\renewcommand{\algorithmicensure}{\textbf{Output:}}
	\caption{Pseudo Code of Our Framework} 
	\label{alg: ours} 
	\begin{algorithmic}
		\REQUIRE Settings for training policy net: training epochs $T_{p}$, batch size $t_{p}$ and optimizer A; Settings for training reliability estimator: dataset size $t_d$, perturb probabilities $p_a$ and $p_b$, training epochs $T_e$, batch size $N_b$ and optimizer B.
	    \STATE Randomly initialize the network parameters $\theta$, $\phi$.
        \STATE // Stage I: training policy net
		\FOR{$T=1$ to $T_{p}$}
        \STATE // Interacting with the environment 
		\FOR{$t=1$ to $t_{p}$}
		\STATE Reset the environment and get observations $o_i^t$ for each agent if necessary; 
		\STATE Each agent updates hidden states $h_i^t$, generates and broadcasts messages $m_i^t$, receives messages from others, and chooses actions $a_i^t$;
		\STATE For each agent, compute value function $v_i^t=V_{\phi}(h^t_i,o_i^t)$;
		\STATE Each agent executes action $a_i^t$ and receives reward $r_i^t$ as well as next observation $o_i^{t+1}$ from the environment;
		\STATE Store $(o^i_t, m_i^t, v_i^t, a_i^t, r_i^t, o^{t+1}_i)$ into the replay buffer;
		\ENDFOR 
        \STATE // Updating neural networks 
        \STATE For each transition in the replay buffer, compute $Q$ value according to (\ref{eqa:qcompute});
	\STATE Calculate $L_V(\phi)$ and corresponding gradients $\nabla_{\phi}L_V(\phi)$ based on the transitions in the replay buffer according to (\ref{eqa:value_loss}); Calculate $\nabla_{\theta}J(\theta)$ according to (\ref{eqa:gradient});
		\STATE Update $\theta$ and $\phi$ using $\nabla_{\theta}J(\theta)$, $\nabla_{\phi}L_V(\phi)$, and optimizer A;
		\ENDFOR 
        \STATE // Stage II: generating dataset
        \FOR{$t=1$ to $t_d$}
        \STATE Reset the environment and get observations $o_i^t$ for each agent if necessary;
        \STATE Each agent updates hidden states $h_i^t$, generates and broadcasts messages $m_i^t$, receives messages from others, and chooses actions $a_i^t$;
        \STATE Replace messages with probability $p_a$. If a message is replaced, then it has a probability $p_b$ to be replaced by a random message and a probability $1-p_b$ to be replaced by an adversarial message. Then, all messages, including unperturbed and perturbed ones, are labeled according to their preference for the best action. 
        \STATE Store $x_{ij}=[h_i,o_i,m_j]$ and $y_{ij}$ into a dataset $D$
        \STATE Each agent executes action $a_i^t$ and receives reward $r_i^t$ as well as next observation $o_i^{t+1}$ from the environment;
        \ENDFOR
        \STATE // Stage III: training reliability estimator
        \FOR{$T=1$ to $T_e$}
        \STATE Randomly divide dataset $D$ into batches of size $N_b$, and assume there are $t_{e}$ batches in total.
        \FOR{$t=1$ to $t_e$}
        \STATE Use the $t$-th batch to calculate $L_R(\psi)$ with (\ref{eqa:sl})
        \STATE Update $\psi$ with $\nabla_{\psi}L_R(\psi)$ as well as optimizer B.
        \ENDFOR
        \ENDFOR
	\end{algorithmic} 
\end{algorithm}

\subsection{Training settings}
\paragraph{Neural Network architectures} An agent uses a GRU module to obtain hidden states, and all other neural networks are fully connected with hidden layer size $128$ and activation function ReLU. Specifically, observations are linearly transformed into vectors of length $128$ before being used. The base action generation module $f_{BP}$ and the message-observation process module $f_{MP}$ both have one hidden layer. The reliability estimator $f_R$ has two hidden layers. 

\paragraph{Training hyperparameters} Training epochs $T_p=2000$, $T_e=100$. Batch sizes $t_p=6000$, $N_b=64$. The training dataset for the estimator contains around $300000$ samples. Perturb probabilities $p_a=0.5$, $p_b=0.5$. We use Adam optimizer with learning rate $0.0003$ during training.

\section{C. Additional Experimental Results}
\begin{figure}[h!]
    \centering
    \begin{minipage}[t]{0.32\textwidth}
    \centering
    \includegraphics[width=1\textwidth]{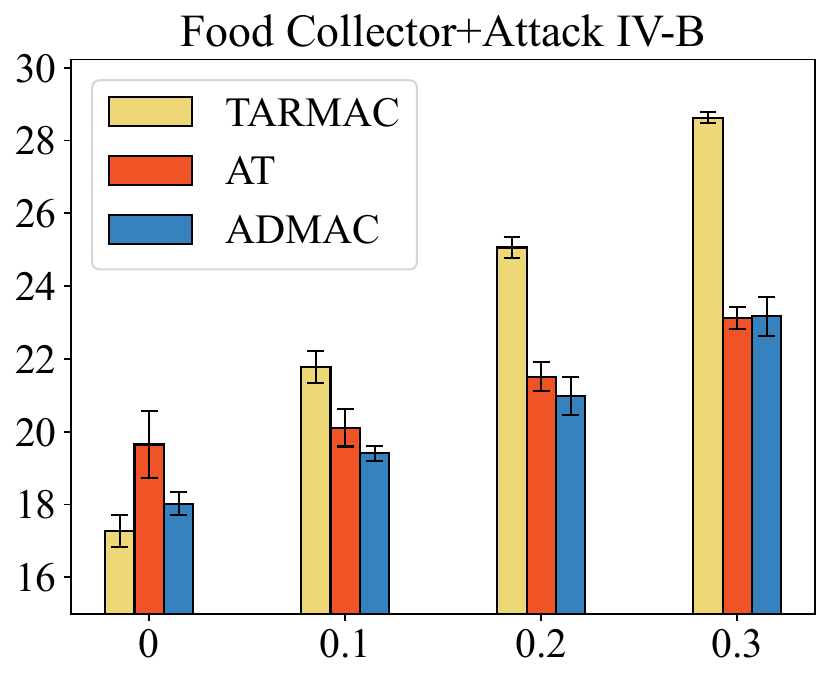}
    \end{minipage}
    \begin{minipage}[t]{0.32\textwidth}
    \centering
    \includegraphics[width=1\textwidth]{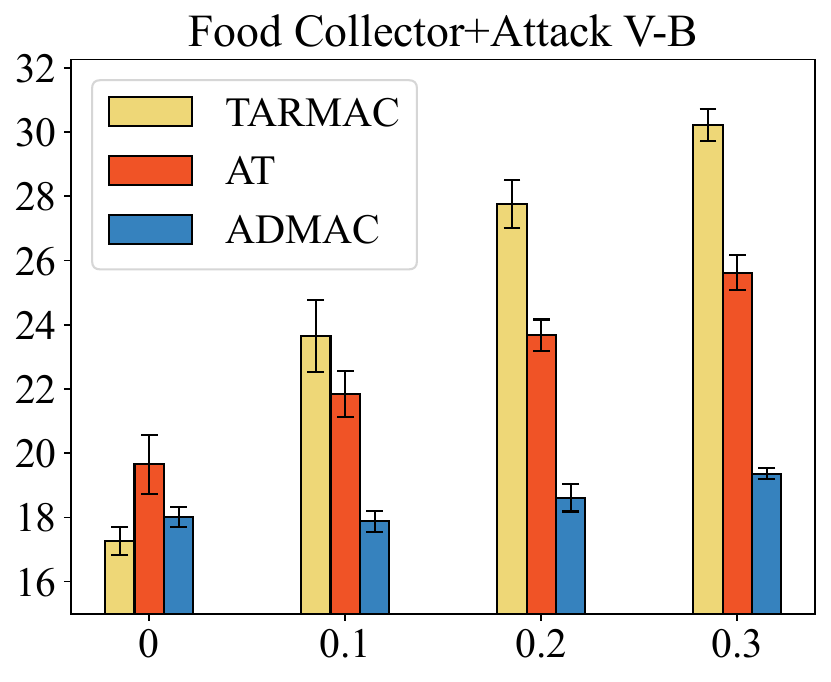}
    \end{minipage}
    \begin{minipage}[t]{0.32\textwidth}
    \centering
    \includegraphics[width=1\textwidth]{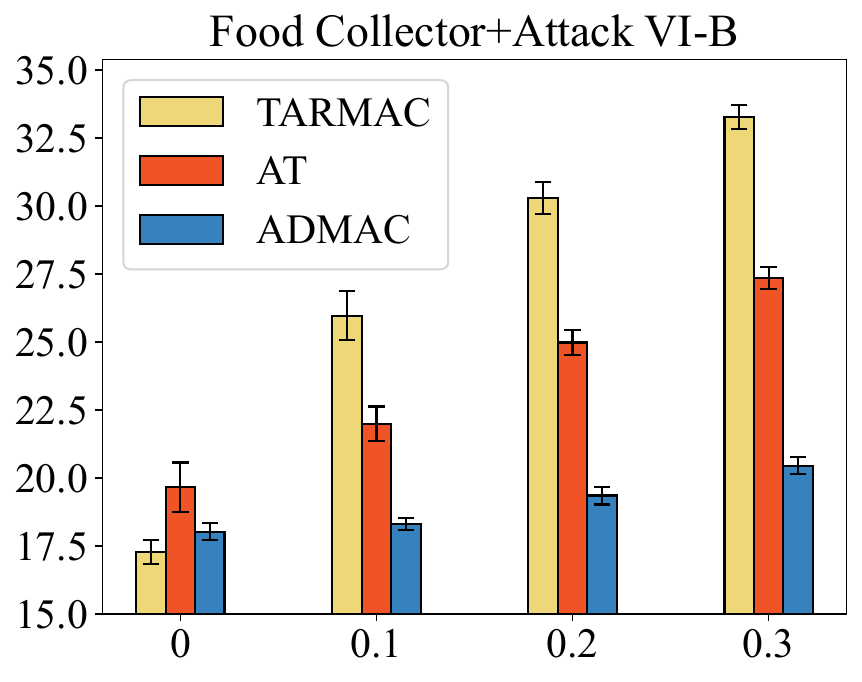}
    \end{minipage}
    
    \begin{minipage}[t]{0.32\textwidth}
    \centering
    \includegraphics[width=1\textwidth]{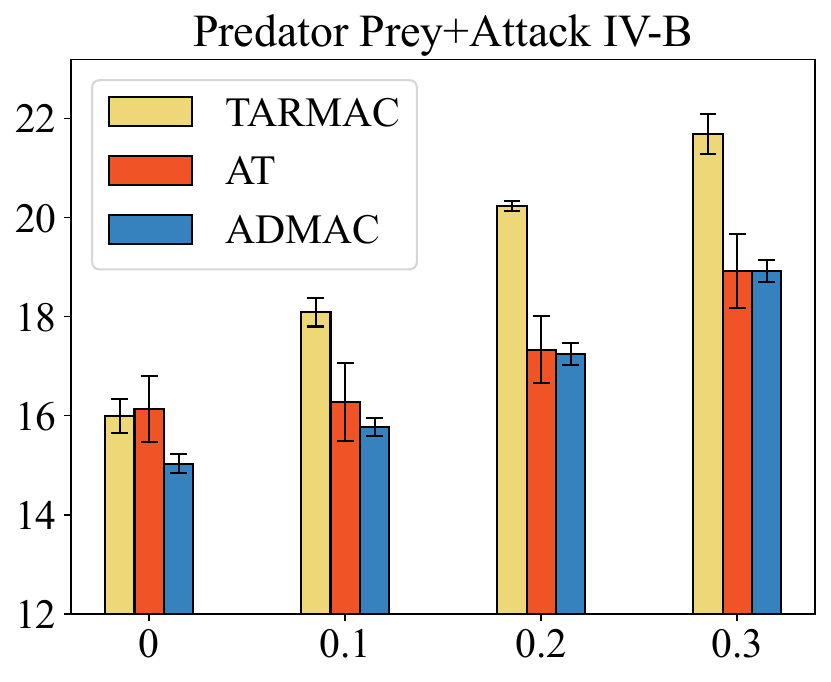}
    \end{minipage}
    \begin{minipage}[t]{0.32\textwidth}
    \centering
    \includegraphics[width=1\textwidth]{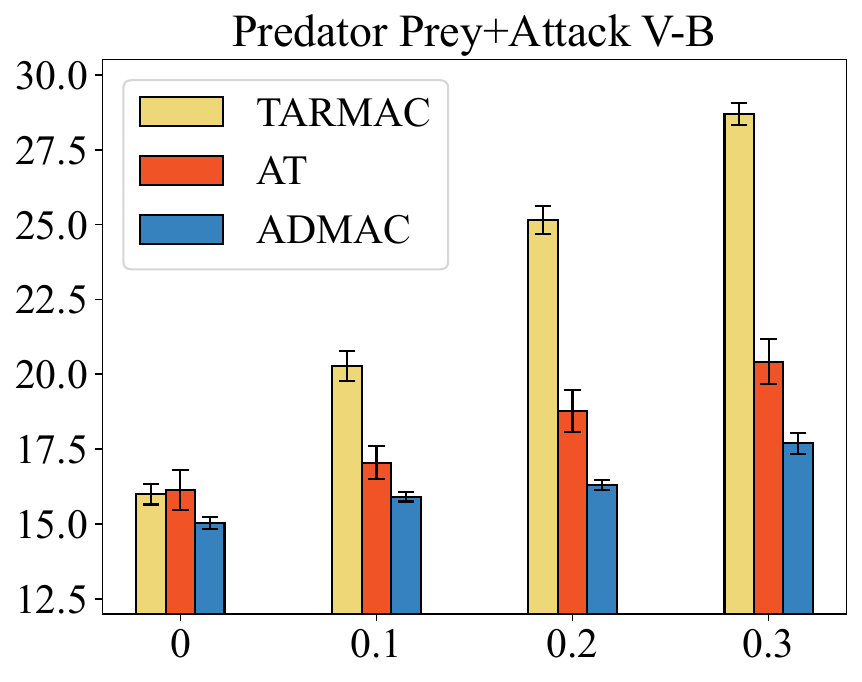}
    \end{minipage}
    \begin{minipage}[t]{0.32\textwidth}
    \centering
    \includegraphics[width=1\textwidth]{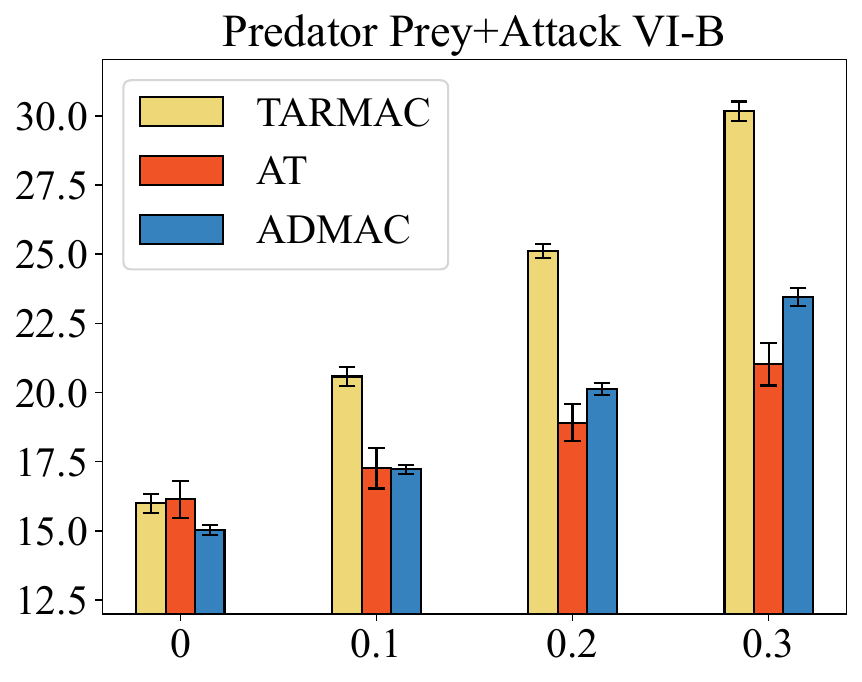}
    \end{minipage}
    
    \begin{minipage}[t]{0.32\textwidth}
    \centering
    \includegraphics[width=1\textwidth]{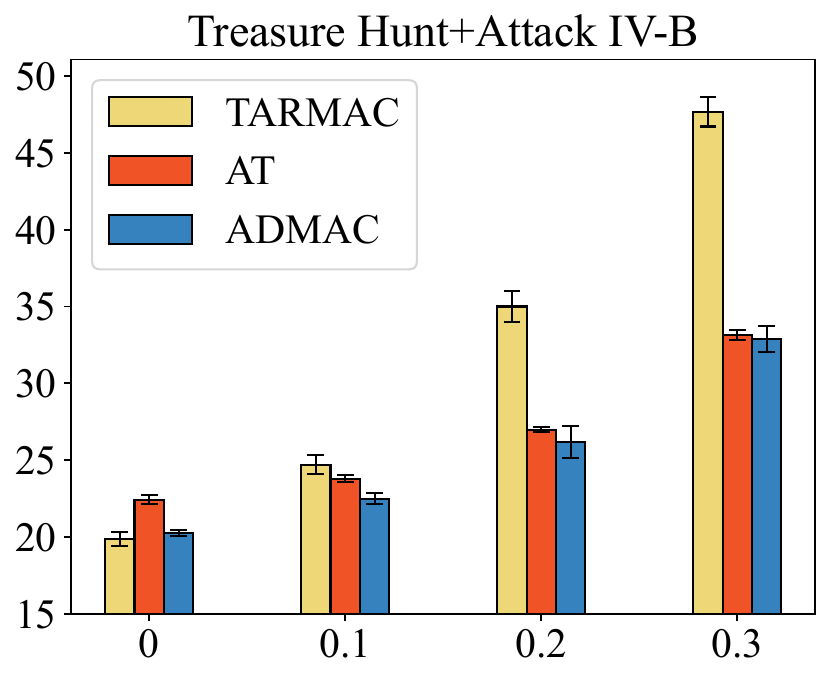}
    \end{minipage}
    \begin{minipage}[t]{0.32\textwidth}
    \centering
    \includegraphics[width=1\textwidth]{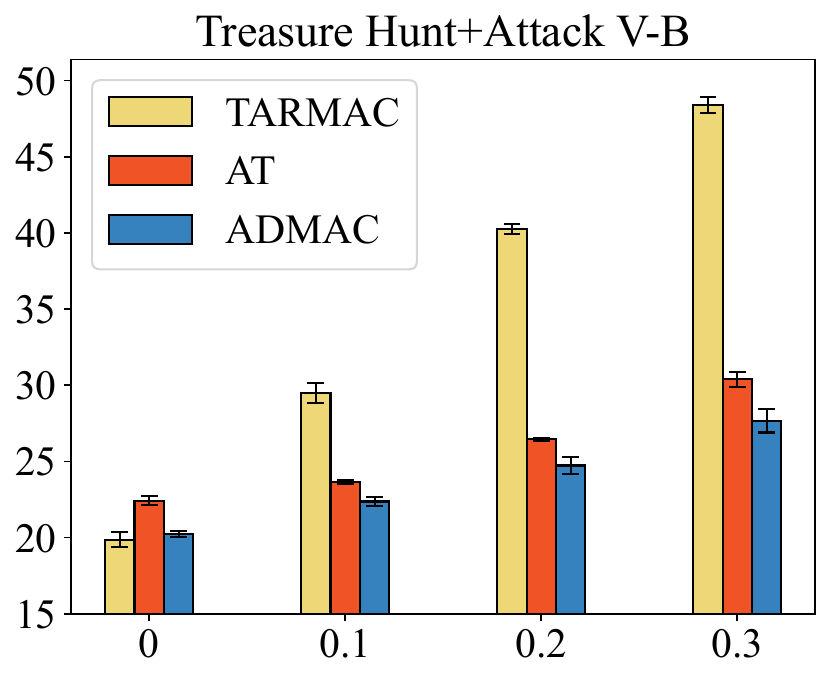}
    \end{minipage}
    \begin{minipage}[t]{0.32\textwidth}
    \centering
    \includegraphics[width=1\textwidth]{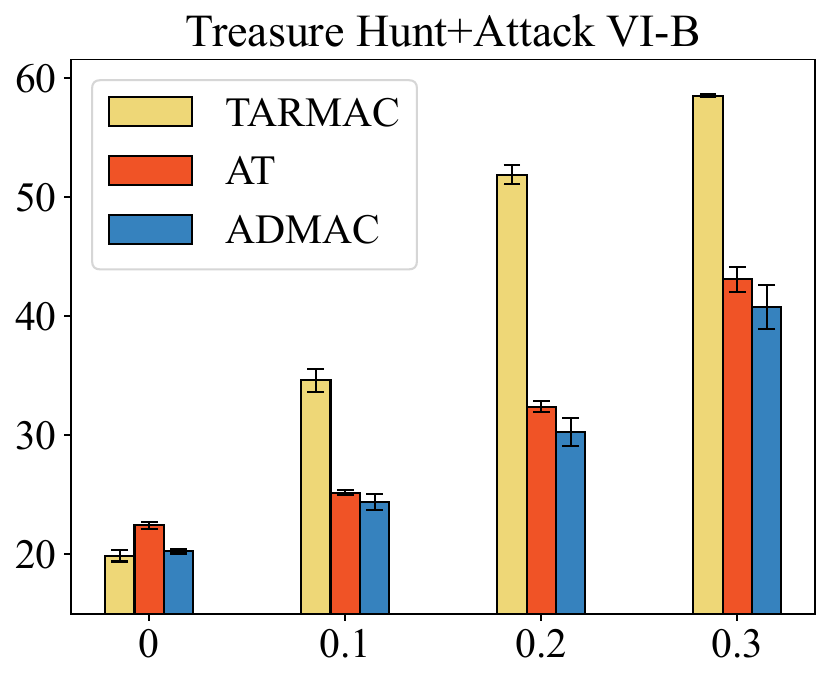}
    \end{minipage}
    
    \caption{Performance of different models under adversarial attacks. }
    \label{fig:test}
\end{figure}

In our experiments, the implemented adversarial attacks have two attack goals:
\begin{equation}
f_A(m_j^t) = \arg\min \sum_{i\neq j} \hat{P}(a^t_{i,best}).
\end{equation}
and
\begin{equation}
f_B(m_j^t) = \arg\max \sum_{i\neq j} D_{KL}(\hat{a}^t_i|a^t_i),
\end{equation}
Since $f_B(m_j^t)$ is not applicable to AME, we only present type-A adversarial attacks in the main manuscript. Fig.\ref{fig:test} presents the performance of models under type-B adversarial attacks. Consistent with the primary experimental findings, our approach exhibits excellent defensive performance.

\end{document}